\documentclass[smallextended]{svjour3}     
\smartqed  
\usepackage{graphicx}
\newcommand{\be}{\begin{equation}}
\newcommand{\ee}{\end{equation}}
\newcommand{\bea}{\begin{eqnarray}}
\newcommand{\eea}{\end{eqnarray}}
\newcommand{\nn}{\nonumber}
\def\xL{x_{\rm L}}
\def\rs{r_{\rm S}}
\def\xs{x_{\rm S}}
\def\ys{y_{\rm S}}

\def\dL{d_{\rm L}}
\def\rL{r_{\rm L}}
\def\L{\Lambda}

\def\vt{\vartheta}

\newcommand{\de}{\hbox{\rm{d}}}

\newcommand{\bb}{\begin{eqnarray}}
\newcommand{\eee}{\nonumber\end{eqnarray}}

\newcommand{\beq}{\begin{equation}}
\newcommand{\eeq}{\end{equation}}

\newcommand{\ba}{\begin{eqnarray}}
\newcommand{\ea}{\end{eqnarray}}
\journalname{General Relativity and Gravitation}
\begin{document}
\title{The Relevance of the Cosmological Constant for Lensing
}
%
\author{Mustapha Ishak\and
        Wolfgang Rindler 
}
%
%
\institute{Department of Physics, The University of Texas at Dallas, Richardson, TX 75083, USA \\
              Tel.: 1-972-883-2815\\
              Fax:  1-972-883-2848\\
              \email{mishak@utdallas.edu}\\
	      \email{rindler@utdallas.edu}
           }
\date{Received: 26 November 2009 / Accepted: 14 March 2010}
%
\maketitle
\begin{abstract}
This review surveys some recent developments concerning
the effect of the cosmological constant on the bending of 
light by a spherical mass in Kottler (Schwarzchild-de 
Sitter) spacetime. Some proposals of how such an effect 
may be put into a setting of gravitational lensing in 
cosmology are also discussed. The picture that emerges 
from this review is that it seems fair to assert that the 
contribution of $\Lambda$ to the bending of light has by 
now been well established, while putting the $\Lambda$ 
light-bending terms into a cosmological context is still 
subject to some interpretation and requires further work 
and clarification.
\keywords{bending of light \and lensing \and cosmological constant}
%
\end{abstract}
%
\section{Introduction}
\label{intro}
%
In view of the present candidacy of the cosmological constant  $\Lambda$ as a possible cause of the observed acceleration of the universe, any non-cosmological manifestation of $\Lambda$ could be of interest. Already in his 1923 treatise "The Mathematical Theory of Relativity" [Sec.45 in \cite{Eddington}]. A. S. Eddington examined one possible such manifestation, namely the contribution of $\Lambda$ to the perihelion advance of the planets. Soon thereafter he applied his results to the special case of Mercury, and found that if $\Lambda$ were greater than $5\times 10^{-42} cm^{-2}$ it would increase the centennial advance of the perihelion from $43^{\prime\prime}$ to more than $44^{\prime\prime}$, which would have been detected by the astronomers, see also \cite{Rindler1}. Eddington never seems to have made a similar calculation for the other then-"crucial" effect of GR, the bending of light. Possibly in view of Einstein's later renunciation of the $\Lambda$-term, interest in this matter faded. In any case, when compared to the cosmologically determined value of about $10^{-56}cm^{-2}$, Eddington's upper bound strongly suggests that a "locally" detectable $\Lambda$-effect could only be expected on cosmologically significant length scales. 

In modern times the problem was taken up again – apparently for the first – time in a much-quoted 1983 paper by N. J. Islam \cite{Islam}. There, the author re-derived Eddington's $\Lambda$ limit of $10^{-42}cm^{-2}$ from the perihelion advance of Mercury; but, more importantly, he comes to the conclusion that $\Lambda$ has no influence on the bending of light. The case for this noninfluence of $\Lambda$ was then remade and reaffirmed by various other authors. N. J. Islam's argument was very simple: In the Kottler metric \cite{Kottler}, which is the modified version of the Schwarzschild metric when $\Lambda$ is included in the field equations, $\Lambda$ modifies the orbits of all massive particles, but drops out of the coordinate orbital equation for photons.

However, in a 2007 paper by the present authors \cite{RindlerandIshak} it was pointed out that the coordinate differential equation for the orbit and its integral are only half the story.  The other half is the metric itself.  It is only the metric that determines the geometric and measurable properties of the coordinate orbital equations.  When that is taken into account, a quite different picture emerges:  $\Lambda$ does indeed contribute to the bending of light.

Apart from its fundamental aspect, this matter is of potential interest to the many people involved in gravitational lensing where the approach to the problem is generally somewhat different from the basic geometrical one of \cite{RindlerandIshak}. Nevertheless, most authors seem to have accepted the existence of $\Lambda$-bending in principle, though some have expressed reservations, and others have obtained, by different methods, somewhat different formulae as we will discuss.

\section{Light bending in Kottler (Schwarzschild-de Sitter) spacetime}
%
\subsection{The contribution of the cosmological constant to the bending of light}
%
As is well known, the Schwarzschild metric is the unique solution (up to coordinate transformations) of Einstein's original vacuum field equations, when applied to the vacuum between a spherically symmetric central massive object and a possibly surrounding concentric mass distribution (or simply an infinite vacuum). By Birkhoff's theorem \cite{Bonnor}, any purely radial and spherically symmetric motion of the central object or the surrounding masses has no effect on the staticity of the intervening vacuum. 

When Einstein's equations are extended by the so-called cosmological or $\Lambda$-term, the unique successor to the Schwarzschild metric was first obtained by Kottler \cite{Kottler}:
\begin{equation}
ds^2=f(r) dt^2 - f(r)^{-1} dr^2-r^2 (d\theta^2+sin^2(\theta) d\phi^2),
\label{eq:metric}
\end{equation}
where
\begin{equation}
f(r) \equiv 1-\frac{2m}{r}-\frac{\Lambda r^2}{3}
\label{eq:alpha}
\end{equation}
and where units are chosen so as to make $c=G=1$, as will be done throughout this paper unless otherwise stated; $m$ is then the mass of the central object expressed in length units, the only units remaining. When $\Lambda=0$, Kottler's metric reduces to Schwarzschild's, and when $m=0$ it reduces to the static version of de Sitter space. These limits are also approximated by the Kottler metric when $r$ is small and when $r$ is large, respectively. By an extension of Birkhoff's theorem \cite{Bonnor}, any purely radial and spherically symmetric motion of the masses involved leaves the staticity of the Kottler metric unchanged. 
\begin{figure}
\begin{center}
\includegraphics[width=3.5in,height=2.5in,angle=0]{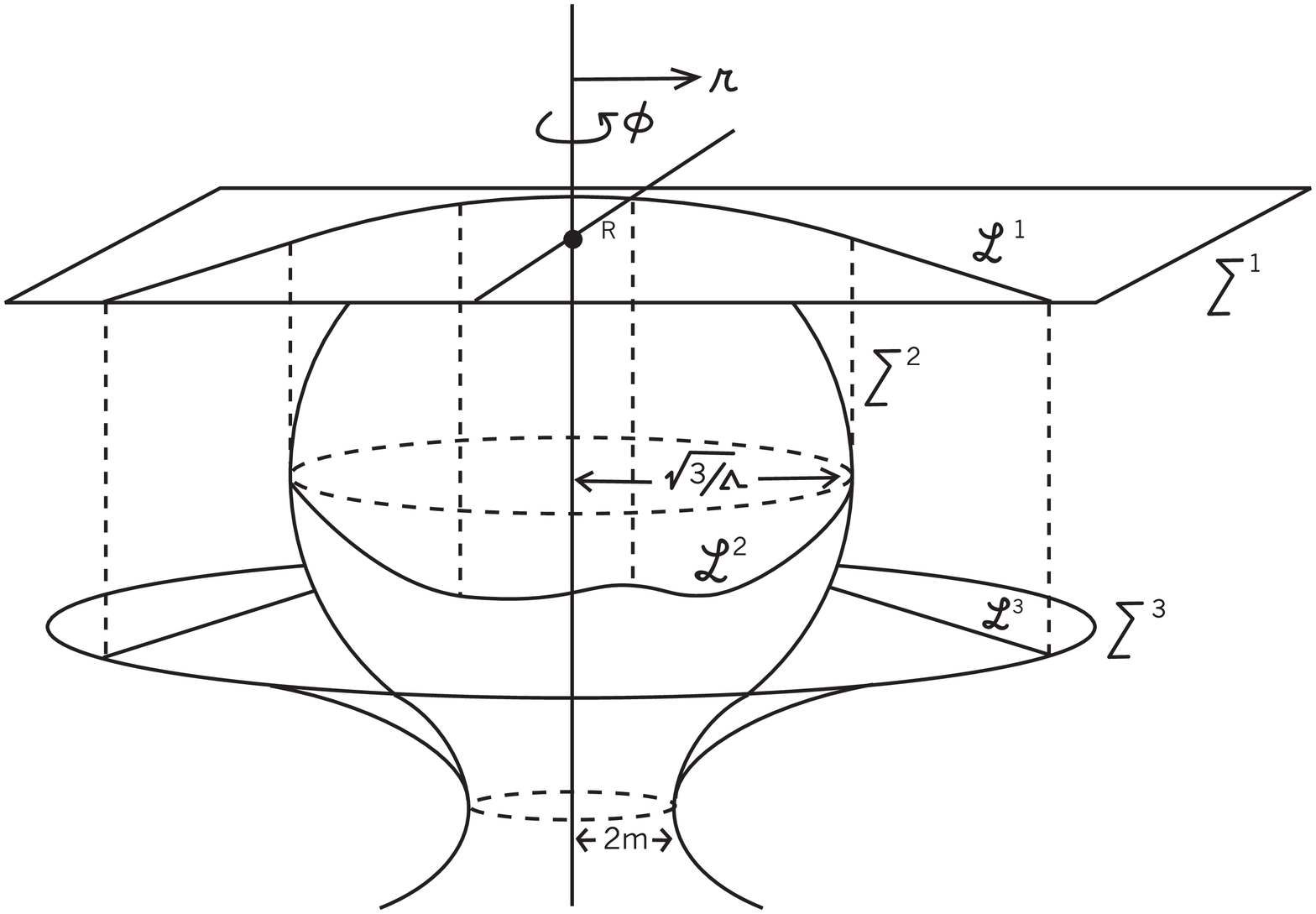}
\caption{\label{fig:figure1} 
Schwarzschild and Schwarzschild-de Sitter geometries.  $\Sigma^3$ is the Flamm paraboloid representation of a central coordinate plane in Schwarzschild;  $\Sigma^2$ is the corresponding surface in Schwarzschild-de Sitter;  $\Sigma^1$ is an auxiliary plane with an $r,\phi$ graph, $\mathcal{L}^1$, of the orbit equation (\ref{eq:solution}).  The curves $\mathcal{L}^2$ and $\mathcal{L}^3$ are the vertical projections of $\mathcal{L}^1$ onto $\Sigma^2$ and $\Sigma^3$, and represent the true spatial curvature of the orbits.
 }
\end{center}
\end{figure}

The geometric modification that Kottler brought to Schwarzschild is highly significant for our purposes. Because of the spherical symmetry of the spatial part of both these metrics, it is clear that any orbit of a particle or a photon must lie entirely in one of the central coordinate "planes" typified by the equatorial plane $\theta=\pi/2$. In fact all orbit calculations are usually made in this representative plane. Of course, although the locus $\theta=\pi/2$ is plane in the coordinate sense, it is \textit{not} plane when measured with rulers. The ruler geometry of this locus is determined by the spatial part of the metric (\ref{eq:metric}), which for the Schwarzschild case ($\Lambda=0$) is 
\begin{equation}
dl^2=\Big{(}1-\frac{2m}{r}\Big{)}^{-1}dr^2+r^2 d\phi^2,
\label{eq:S2metric}
\end{equation}
and for the de Sitter case ($m=0$) is 
\begin{equation}
dl^2=\Big{(}1-\frac{\Lambda r^2}{3}\Big{)}^{-1}dr^2+r^2 d\phi^2.
\label{eq:SdS2metric}
\end{equation}
As is well known, (\ref{eq:S2metric}) has the intrinsic geometry of the upper half of the so-called Flamm Paraboloid \cite{Flamm} which, when displayed as a surface of revolution in ordinary Euclidean space, has the equation 
\begin{equation}
z^2=8 m (r-2m),
\end{equation}
$z$ being the distance along the axis and $r$ the distance from the axis. The corresponding intrinsic geometry of the metric (\ref{eq:SdS2metric}) is that of the lower half of a 2-sphere of radius $\sqrt{3/\Lambda}$, with the origin at the southpole, $\phi$ being longitude and $r$ again the distance from the axis. In the Schwarzschild case the range of $r$ (region of staticity) is bounded below by $r=2m$, the waist circle of the paraboloid. In the de Sitter case it is bounded above by $r=\sqrt{3/\Lambda}$, the equator of the sphere.

The spatial geometry of the Kottler metric (\ref{eq:metric}), being essentially a combination of (\ref{eq:S2metric}) and (\ref{eq:SdS2metric}), is illustrated in Fig.1. For small $r$ it approximates the Flamm funnel down to $r\approx 2m$, while for large $r$ it widens into an approximate de Sitter sphere up to $r\approx \sqrt{3/\Lambda}$. The precise $r$-values of these boundaries are given by the two positive zeros of $f(r)$. These correspond to event horizons, but will not concern us here, since the region of interest lies well in between.
\begin{figure}
\begin{center}
\includegraphics[width=3.5in,height=1.5in,angle=0]{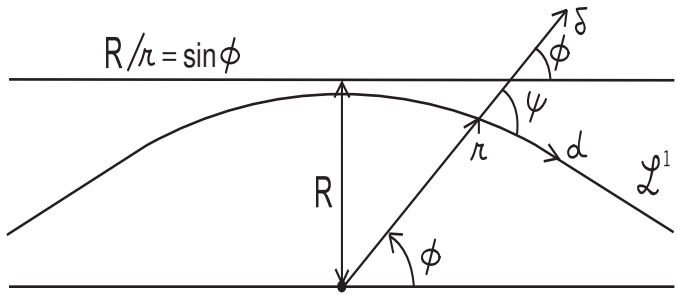}
\caption{\label{fig:figure2} 
The orbital map.  This is a plane graph of the orbit equation (\ref{eq:solution}) and coincides with $\Sigma^1$ in Figure 1. The one-sided deflection angle is $\psi-\phi\equiv\epsilon$.
}
\end{center}
\end{figure}

The work of \cite{RindlerandIshak} is based on the well-known first approximation orbital equation for photons in the $\theta=\pi/2$ plane of the Schwarzchild metric 
\begin{equation}
\frac{1}{r}=u=\frac{sin(\phi)}{R}+\frac{3m}{2R^2}\Big{(}1+\frac{1}{3}cos(2\phi)\Big{)},
\label{eq:solution}
\end{equation}
which, as mentioned earlier, applies equally in the Kottler metric. The orbits are parameterized by $R$, the distance of the zeroth-order (straight-line) approximative solution from the center (see Fig.2). The relation of $R$ to the physically meaningful coordinate distance (area distance) $r_0$ of closest approach is obtained by setting $\phi=\pi/2$ in (\ref{eq:solution}): 
\begin{equation}
\frac{1}{r_{0}}=\frac{1}{R}+\frac{m}{R^2}.
\label{eq:r_0R}
\end{equation}
Now, in Schwarzschild space it is easy to obtain the total deflection angle suffered by a ray of light coming from infinity and going to infinity by letting $r \longrightarrow \infty$ in (\ref{eq:solution}). At infinity, the coordinate plane $\theta=\pi/2$ is essentially Euclidean, and the coordinate angle $\phi$ is also the measured angle (see Fig.2). Not so in Kottler space. There $r \longrightarrow \infty$ is impossible because of the spherical nature of the geometry (see Fig.1). The approach adopted in \cite{RindlerandIshak} is to find the angle $\Psi$ \textit{measured} by an observer at an arbitrary point $(r,\phi)$, between a ray coming radially from the center and the ray skirting the lens (see Fig.2). This angle is given by the standard formula
\begin{equation}
cos (\psi)= {\frac{g_{ij} d{^i} \delta^{j}}{(g_{ij} d^{i} d^{j})^{1/2}(g_{ij}\delta^{i} \delta^{j})^{1/2}}}
\label{eq:cosine}
\end{equation}
for the angle between $d^i$ and $\delta^i$. It is precisely here where $\Lambda$ comes into play, via the metric coefficients $g_{ij}$ of the Kottler metric (\ref{eq:metric}). 

For an $r$-value sufficiently far from the lens for $\phi$ and $\psi$ to be small (see Figure 2), the results of \cite{RindlerandIshak} (especially Eqs. (13) and (16)) can be shown (by a sequence of Taylor approximations, working to first order in $\phi$, $\psi$, $m/R$ and $\Lambda r^2$) to be equivalen to 
\be
\psi= \phi +\frac{2m}{R}- \frac{\Lambda R^3}{6(2m+\phi R)}.
\label{eq:bendingangle}
\ee
This would seem to definitely indicate a contribution of $\Lambda$ to the bending of light. [The term $\frac{2m^2}{R^2}$ in the corresponding Eq. (17) of Ref.  \cite{RindlerandIshak} is superfluous and innacurate since the expansion was not done to the second order but see equation (\ref{eq:result2}) below] As can be seen from Fig.2, the one-sided actual deflection angle is $\psi-\phi$. Thus, in the Schwarzschild case where $r  \longrightarrow \infty$ makes sense, $r  \longrightarrow \infty$ implies $\psi \longrightarrow 0$ and then (\ref{eq:bendingangle}) gives the familiar Einstein deflection.

A particularly transparent application of Eq.(\ref{eq:bendingangle}) is given in \cite{IRD}. It concerns an Einstein ring, a well-known example of gravitational lensing, where a concentrated source directly behind the lens is seen by an observer directly in front of the lens apparently spread into a ring. The case most easily dealt with by formula (\ref{eq:bendingangle}) is when the source and observer are equidistant from the lens, so that in Fig.2 the observer is at $\phi=0$ and consequently,  by (\ref{eq:bendingangle}),
\be
\psi=\frac{2m}{R}-\frac{\Lambda R^3}{12 m}.
\label{eq:bendingangle2}
\ee
This, in principle, appears to offer a method of measuring the $\Lambda$-contribution if we know $m$ and $r_0$, the radius of the lens -- which is related to $R$ by Eq. (\ref{eq:r_0R}) . 

Of course, lensing on the cosmological scale is not done by an observer at rest relative to the source. But any radial velocity V of the observer would simply increase the entire $\psi$ of Eqs. (\ref{eq:bendingangle}) and (\ref{eq:bendingangle2}) by the usual aberration factor $\sqrt{(1+V)/(1-V)}$ of special relativity, and thus leave the {\textit{relative}} contribution of $\Lambda$ unchanged. 
Neverthless, in a universe with cosmological constant $\Lambda$, the velocity $V$ is $\Lambda$-dependent, and the cosmological expansion effect of $\Lambda$ (via aberration) can exceed its above discussed geometric effect by many orders or magnitude. For example, in a simple de Sitter universe, where $V\approx Hr=\sqrt{\Lambda/3}r$, the abberation factor is $\sim(1+\sqrt{\Lambda/3}r)$ and then the dominant $\Lambda$-term in (\ref{eq:bendingangle2}) becomes
\be
\frac{2mr}{R}\sqrt{\frac{\Lambda}{3}}=R\sqrt{\frac{\Lambda}{3}}=\sqrt{\frac{2mr\Lambda}{3}},
\label{dominantterm}
\ee
where we have used Eq. (\ref{eq:solution}) with $\phi=0$ to relate $r$, $R$, and m. Since $\sqrt{\Lambda}$ is very much bigger. Since  $\Lambda$ is small, root $\Lambda$ is much bigger, and so the aberration effect of $\Lambda$ overshadows its geometric effect.
%
\subsection{Higher order mass terms}
%
It is of interest to compare the $\Lambda$-terms to higher-order mass terms, for which purpose we must extend the calculation above to higher orders. Proceeding in the usual way, see for example \cite{Bodenner,IshaketalMNRAS}, one writes
\begin{equation}
u=u_0 [sin(\phi) + (m u_0) \delta u_1 + (m u_0)^2 \delta u_2]
\label{eq:SecondOrderTrial} 
\end{equation}
where $u \equiv \frac{1}{r}$ and $u_0 \equiv \frac{1}{R}$. Substituting this into the null orbit ODE, e.g. \cite{Rindler,MTW}  
\begin{equation}
\frac{d^2u}{d\phi^2}+u=3 m u^2, \,\,\,\, (u \equiv 1/r).
\label{eq:ODE}
\end{equation} 
 and collecting terms of equal powers of $Mu_0$ gives the following two equations: 
\begin{equation}
\frac{d^2 \delta u_1}{d\phi^2}+\delta u_1 = 3 \sin^2 \phi
\label{eq:ODE1}
\end{equation}
\begin{equation}
\frac{d^2 \delta u_2}{d\phi^2}+\delta u_2 = 6 \delta u_1 \sin \phi.
\label{eq:ODE2}
\end{equation}
Solving (\ref{eq:ODE1}) and (\ref{eq:ODE2}) for $\delta u_1$ and $\delta u_2$ and substituting them into (\ref{eq:SecondOrderTrial}) gives the solution 
\be
\frac{1}{r}=\frac{\sin \phi}{R} + \frac{3 m}{2 R^2}\Big{(}1+\frac{\cos 2 \phi}{3} \Big{)} +  \frac{3 m^2}{16 R^3}\Big{(}10 \pi \cos \phi - 20\phi \cos \phi - \sin 3 \phi\Big{)}.
\label{eq:SecondOrderSolution} 
\ee

Now, one can differentiate (\ref{eq:SecondOrderSolution}) and multiply by $r^2$ to obtain 
\be
\frac{dr}{d\phi}=-\frac{r^2}{R} \cos \phi + \frac{m r^2}{R^2} \sin 2\phi+ \frac{15m^2 r^2}{4 R^3}\Big{(}\cos \phi+\frac{3}{20} \cos 3\phi + (\frac{\pi}{2}-\phi)\sin \phi\Big{)}.
\label{eq:A2}
\ee
After some steps, it follows from (\ref{eq:cosine}) and (\ref{eq:A2}) that the bending angle (at $\phi=0$) to the second-order is given by 
\begin{equation}
\alpha \approx 4 \frac {m}{R}+ \frac{15 \pi}{4}\frac {{m}^{2}}{{R}^{2}} -
\frac {\Lambda{R}^{3}}{6 m}
\label{eq:result2}
\end{equation}
where $\alpha\equiv 2\psi$ of equation (\ref{eq:bendingangle2}). The first and second-order mass terms are the same as in the expansion in terms of the impact parameter $b$, see e.g. \cite{Keeton} for the Schwarzschild case. 
%
%
\subsection{$\Lambda$-contributions from perturbations of the equations of motions}          
%
Sereno \cite{SerenoI} performed calculations based on perturbations of the equations of motion in the weak deflection limit and a term by term integration. He considered the light orbital equation from the source at coordinates $\{r_\mathrm{s}, \phi_\mathrm{s} \}$ to the observer at coordinates $\{r_\mathrm{o}, \phi_\mathrm{o} =0 \}$, written in terms of the first integral of motion $b \equiv \dot{\phi}r^2$ as
\be
\label{geo1}
\phi_{s}=\pm \int \frac{dr}{r^2} \Big{[} \frac{1}{b^2}+\frac{1}{r_\Lambda^2}-\frac{1}{r^2}+\frac{2m}{r^3} \Big{]} ^{-1/2},
\ee
where $r_\Lambda \equiv \sqrt{3/\Lambda}$. Next, he expanded the equation for the point of closest approach in the weak deflection limit to find 
\be
\label{geo2}
r_\mathrm{min} \simeq b  \left\{
1 - \frac{m}{b}   -    \frac{3 m^2}{2 b^2}  -   \frac{4 m^3}{b^3}  - \frac{105 m^4}{8 b^4}   -\frac{b^2}{2r_\L^2} 
 \right\}.
\ee
Then he integrated term-by-term Eq.~(\ref{geo1}) to find
\ba
\phi_\mathrm{s} &  = & -\pi -\frac{4 m}{b} + b \left(\frac{1}{r_\mathrm{s}}+\frac{1}{r_\mathrm{o}}\right) -\frac{15 m^2 \pi }{4 b^2} -\frac{128 m^3}{3 b^3}+\frac{b^3}{6}  \left(\frac{1}{r_\mathrm{s}^3}+\frac{1}{r_\mathrm{o}^3}\right)-\frac{3465 m^4 \pi }{64 b^4}  \\
& - & \frac{3584 m^5}{5 b^5} - \frac{2 m b}{r_{\Lambda }^2}- \frac{ m b^3}{4} \left(\frac{1}{r_\mathrm{s}^4}+\frac{1}{r_\mathrm{o}^4}\right)+  \frac{3 b^5}{40} \left(\frac{1}{r_\mathrm{s}^5}+\frac{1}{r_\mathrm{o}^5}\right)- \frac{b^3}{ 2r_{\Lambda }^2}  \left( \frac{1}{r_\mathrm{s}}+\frac{1}{r_\mathrm{o}}\right)
+  {O}(\epsilon^6). \nonumber  \label{geo3}
\ea
with the contribution of the cosmological constant appearing in the last term and also the term $2 b m/r_\L^2$ that Sereno qualified as local. He then derived for small angles the relation between the constant of motion $b$ and the angle $\vt$ between the tangent to the photon trajectory at the observer abd the radial direction to the lens as measured in the locally flat observer's frame 
\beq
\label{lens2}
\vt \simeq \frac{b}{r_\mathrm{o}} + \frac{b^3}{6 r_\mathrm{o}^3} \left[ 1-\frac{6 m r_\mathrm{o}}{b^2} -\frac{3 r_\mathrm{o}^4}{b^2 r_\L^2}  \right]
\eeq
He then re-wrote the $\Lambda$-contribution in a form that, when evaluated at a point corresponding to $\phi=0$ in Fig. 2 above, reduces to
\be
\delta\vt_{\Lambda}=-\frac{\Lambda b^3}{12m},
\label{eq:equivalent}
\ee
in agreement with Rindler and Ishak \cite{RindlerandIshak}, see Eq. (\ref{eq:bendingangle2}) above 

Sereno also proposed in the same paper a derivation based on the lens equation and the angular diameter distances where he argues that, to  first order, the lens equation obtained has the same $\Lambda$-term as one would obtain from the integration of the geodesic equations and given by (\ref{eq:equivalent}). Sereno concludes that while the cosmological constant contributes to the bending of light, some terms like the term of Ishak and Rindler in Ref. \cite{RindlerandIshak} can be incorporated into the angular diameter distances.  

Finally, in another paper \cite{SerenoII}, Sereno goes back to the term $2 b m/r_\L^2$ of Eq. (\ref{geo3}) and argues that this term describes the coupling between the lens and the cosmological constant and since neither the source's or observer's position enters into it, this term is clearly local. He argues that contrarily to the other $\Lambda$-term, this local term cannot be integrated into the angular diameter distances.  
%
%
\subsection{The $\Lambda$-term derived by passing from coordinate angles to physical angles and null geodesics integration}
%
In a first paper \cite{SchuckerI}, Schucker computed the relation between coordinate bending angles and physical bending angles and found a result with a $\Lambda$ dependence in agreement with reference \cite{RindlerandIshak} as discussed in  section 2.1 above. His calculation of the physical angle measured by a static observer on Earth was based on using a ratio of proper times of flight of photons from the ratio of coordinate times using the Kottler metric. For example, he calculates the physical angle $\alpha_T$ using  \cite{SchuckerI} 
{\small
\be 
\tan \alpha _T=\,\frac{\tilde{\de\tau}}{\de\tau}\,=\,\frac{\tilde{\de t}\sqrt{B(r_T)}}{\de t \sqrt{B(r_T)}}\,=
\,\frac{r_T\,\de\varphi }{\de r/\sqrt{B(r_T)}}\,={\epsilon_T}{\sqrt{B(r_T)}}\sim
{\epsilon_T}{\sqrt{1-\Lambda r_T^2/3}}\sim\alpha _T
\label{eq:Schucker}
\ee
}
where $\epsilon_T$ is the coordinate angle. The square root factor term brings in the $\Lambda$ dependence. 

Schucker did not use any lens equation but used a self-contained method: to integrate the null geodesics in Kottler space; to compute area distances to the lens and source; and finally to compare his results to lensing cluster SDSS J1004+4112, see e.g. \cite{Sharon}. He found, in agreement with \cite{IshaketalMNRAS} that for certain clusters of galaxies, the $\Lambda$ contribution to bending is not negligible. Schucker published a follow up paper on lensing and time delay \cite{SchuckerII} and another paper\cite{SchuckerIII} that we discuss in section 3.3 and where he applied his results to an Einstein-Strauss (Swiss-Cheese) model. 
%
%
\section{$\Lambda$ light-bending in a Kottler (Schwarzschild-de Sitter) vacuole}
%
\subsection{Kottler Vacuole embedded in an FLRW background}
%
The cosmological constant has a very negligible effect on small scales but this is not the case at large scales of distance between galaxies and clusters of galaxies. Therefore, one could measure the term that represent the contribution of the cosmological constant to the bending of light using observations of large Einstein radii where the lens is a galaxy or a cluster of galaxies, and compare them to the first and second order mass terms in the deflection angle.  

The question then is how to model such a system within the cosmological expanding background described by the Friedmann-Lemaitre-Robertson-Walker (FLRW) model. As discussed in for example \cite{Bartelmann}, this is still an open question since we don't have a stable exact solution to Einstein's equations that describes an inhomogeneity (lens) embedded in an FLRW background and proposals or alternative approximations have all some sort of caveat. 

One such proposal is the Einstein-Strauss (Swiss-Cheese) model \cite{Swiss-cheese,Swiss-cheese2,Krasinski} with a non-zero $\Lambda$ that was used in order to study the $\Lambda$ effect on light bending in for example \cite{IshaketalMNRAS,SchuckerIII,Kantowski} and we use in this section as well. The model is known to have an instability to radial perturbations of the boundary of the vacuole, see for example \cite{Krasinski} and references therein, however, our goal here is to find a cut-off location where the $\Lambda$-bending of the lens can be regarded as accomplished. The model is thus used only to estimate the distance to the edge of the Kottler vacuole and use it to give an order-of-magnitude estimate of the $\Lambda$-bending of light term. 

Physically, one is modeling a galaxy or cluster of galaxies and the spacetime surrounding it (up to some boundary with an FLRW spacetime) by a Kottler vacuole. This Kottler vacuole is in the exact theory embedded into the FLRW background using exact matching conditions at the boundary \cite{Darmois,Israel}. One needs to note a point that will have some importance later and that is the size of the lens is much smaller than the size of the vacuole which is determined from the exact matching between the two spacetimes. 
In this setting, the rays of light enter into the vacuole and propagate inside the Kottler spacetime experiencing bending from one boundary of the vacuole to the other and the radius at the boundary, $r_b$, enters into the $\Lambda$-term.  

Mathematically, the Kottler vacuole is exactly embedded into the FLRW spacetime using the spacetime matching formalism of \cite{Darmois,Israel}. The junction relations in this case are simple and well-known in the literature \cite{Swiss-cheese,Swiss-cheese2,Krasinski}, and are given by:
\be
r_{b\,\, in \, Kottler} = a(t) \,\,\textit{r}_{b\,\,in\,FLRW}
\label{eq:cond1}
\ee
and
\be
m_{\,Kottler} = \frac{4 \pi}{3}\,\, r_{b\,\, in \, Kottler}^3\times \rho_{matter\,in\,FLRW}.        
\label{eq:cond2}
\ee
Thus, for a given cluster mass, Eq.(\ref{eq:cond2}) provides a boundary radius where the spacetime transitions from a Kottler spacetime to an FLRW background. One shall assume that all the light-bending occurs in the vacuole according to our previous formulae, and that once the light transitions out of the vacuole and into FLRW spacetime, all bending stops. Noteworthy, the mass-effect falls off quickly whereas the $\Lambda$-effect on the bending of light increases with distance from the source;  so as we mentioned earlier, the point where to cut off the integration, i.e. the boundary of the vacuole, becomes an important factor.   

The results of section 2.1 are applied for the Kottler vacuole here. For the small angle $\phi_b$ at the boundary, equations (\ref{eq:solution}) and  $dr/d\phi\equiv A$ give
\be
u_b=\frac{1}{r_b}=\frac{\phi_b}{R}+\frac{2m}{R^2},\,\,\,\,\,\,\,\,|A|=\frac{r_b^2}{R}\Big{(}1-\frac{2\phi_b m}{R}\Big{)}.
\label{eq:u_boundary}
\ee
Inserting (\ref{eq:u_boundary}) into Eq.(\ref{eq:cosine}) yields  
\be
\psi \approx \tan{\psi} \approx \phi_b + \frac{2m}{R}-\frac{\Lambda \phi_b r_b^2}{6} + \,\,\textrm{higher-order terms}.
\ee
So the bending angle, $\alpha$, is given, to the smallest order in $m/R$ and $\Lambda$, by 
\be
\frac{\alpha}{2}\approx \theta - \phi_b \approx \frac{2m}{R}-\frac{\Lambda \phi_b r_b^2}{6}.
\label{eq:epsilon}
\ee
Now, to the smallest order, Eq. (\ref{eq:u_boundary}) yields $\phi_b=R/r_b$, so one can write from (\ref{eq:epsilon})
\be
\alpha\approx\frac{4m}{R}-\frac{\Lambda R r_b}{3}
\ee
where $R$ is related to the closest approach by Eq. (\ref{eq:r_0R}) and $r_b$ is the boundary radius between vacuole and FLRW, and is determined from Eq. (\ref{eq:cond2}). Next, this result can be generalized to higher-order mass-terms giving 
\be
\alpha \approx 4 \frac {m}{R}+ \frac{15 \pi}{4}\frac {{m}^{2}}{{R}^{2}} + \frac {305}{12} \frac {{m}^{3}}{{R}^{3}} -\frac{\Lambda R r_b}{3}
\label{eq:alpha_final}
\ee
In \cite{IshaketalMNRAS}, the authors used the above setting and some selected systems of distant galaxies or clusters of galaxies that are lenses with large Einstein radii and found that despite the smallness of  $\Lambda$, the Einstein first-order term in the bending angle due to these systems is only by some $10^3$ bigger than the $\Lambda$-term and that for the lens systems they considered (see their Table 1), the contribution of the cosmological constant term is larger than the second-order mass-term, with for example ratios going from 2 to 8 for various cluster and galaxy systems. As we discuss in section 3.4 below, Kantowski et al. \cite{Kantowski} also used the Einstein-Strauss model but found a different $\Lambda$-term.
%
%
\subsection{An upper bound on the value of Cosmological Constant from Light-Bending by clusters of galaxies}
%
While current best constraints on the value of $\Lambda$ come from cosmology
 with $\Lambda \sim 10^{-56} cm^{-2}$, see e.g. \cite{RindlerLambda,obs1,obs2,obs3,obs4,obs5,obs6,obs7,obs8,obs9,obs10}), it
has always been of interest to find upper bounds on the value of $\Lambda$
from other astrophysical considerations. These include upper
bounds from planetary precessions of the order of $\Lambda \leq
\,\,10^{-46} cm^{-2}$ see for example \cite{Sereno,Kagramanova} and references therein. 
The argument used in these upper-bound estimates is that the 
contribution of $\Lambda$ cannot exceed the uncertainty in the targeted measurement. 

As was discussed in \cite{IshaketalMNRAS} and the references therein, the uncertainty in the measurements of the bending angle is around $\Delta \alpha \sim$ 5-10$\%$ for several of the systems considered in their Table 1. Thus, if the contribution of $\Lambda$ cannot exceed the uncertainty in the bending angle for these system, then it follows that 
\begin{equation}
\Lambda \leq \frac{3\, \Delta \alpha}{R\,\, r_b}.  
\end{equation}
For example, with $\Delta \alpha = 10\%$, one finds from the system Abell 2744 \cite{Smail1991,Allen} that   
\be
\Lambda \leq  4.23\,\,\,10^{-54} cm^{-2}.
\label{eq:limit}
\ee
Interestingly, these limits are the best observational upper bound on the value of $\Lambda$ after cosmological constraints and are only two orders of magnitude away from the value determined from cosmological constraints. 
This result provides an improvement of 8 orders of magnitude on previous upper bounds on $\Lambda$ that were based on planetary or stellar systems, e.g. \cite{Sereno,Kagramanova}.

In fact, the limit (\ref{eq:limit}) should be considered as a rough estimate only because $\Lambda$ also enters into the expression of the angular diameter distances and \cite{IshaketalMNRAS} noted that this estimation can be affected by a factor of two or less when the uncertainty on $\Lambda$ from the angular diameter distances is accounted for. Furthermore, there is room for improving this upper bound calculation by doing a thorough error analysis where all the other parameters and factors that affect lensing along with their uncertainties are measured independently from lensing and included in the analysis. 
%
\subsection{Piece-wise integration of null geodesics in an Einstein-Strauss model}
%
Following on his previous papers on $\Lambda$ lensing \cite{SchuckerI}, Schucker performed piece-wise integrations of null geodesics \cite{SchuckerIII} in the Einstein-Strauss model with a cosmological constant. He calculated deflection angles following his approach and Eq. (\ref{eq:Schucker}) above, taking into account the recession velocity of an observer moving with the Hubble flow of the background spacetime. As noted in his paper, Schucker agreed with the results of \cite{RindlerandIshak}, that a positive cosmological constant decreases the bending of light by an isolated spherical mass but found that the $\Lambda$-contribution is attenuated by a homogeneous FLRW mass distribution around the mass and by the recession of the observer. As he explained there, similar screening was also partially obtained in the work of \cite{IshaketalMNRAS} but his work using a piecewise integration of geodesics and various observers allows one to quantify this screening. 

Schucker then proceeded with a comparison of the theoretical results and lensed quasar system SDSS J1004+4112 \cite{Sharon}. He discussed the dependence of the contribution of $\Lambda$ to light-bending on estimates of the cluster mass and the recession velocities assumed for the observer. Using a Kottler spacetime, he found that a 20$\%$ increase of $\Lambda$ increases the cluster mass (lens) by 20$\%$ for the observer at rest and by 10$\%$ for the comoving observer (moving with velocity $v=H_0 r$). He then redid the calculations in an Einstein-Strauss universe with a Kottler vacuole matched inside an FLRW background. He found that a 20$\%$ increase of $\Lambda$ decreases the cluster mass by 5$\%$ for the comoving observer in the FLRW background, noting that the $\Lambda$ effect produces more attenuation. Schucker concluded that a homogeneous Friedmann background when added around the spherical mass and the recession of the observer diminishes the effect of $\Lambda$ without however canceling it \cite{SchuckerIII}. 

Schucker also points out at the end of his paper \cite{SchuckerIII} that the $\Lambda$ bending terms can have an interpretation as physical velocity of the observer from the lens. He also states there the following: ``In their original paper \cite{RindlerandIshak}, Rindler and Ishak state explicitly that the observer is at rest with respect to the central mass. The same assumption is made in the earlier literature claiming that the bending of light were independent of $\Lambda$. In my view, this controversy is now settled in favor of Rindler and Ishak \cite{RindlerandIshak}". 
%
\subsection{Lensing corrections due to $\Lambda$ in Flat $\Lambda$CDM model}
%
In a recent paper by Kantowski, Chen and Dai \cite{Kantowski} the aim is to reexamine all higher order corrections to the Einstein bending angle, not just that of $\Lambda$. Once again, an exact Swiss-Cheese solution with Kottler in the vacuole and flat FLRW around it is used. Angles are measured by comoving Friedmann observers. Our interest in this paper is chiefly in what it concludes about the $\Lambda$-correction. Kantowski et al. take careful account of the fact that the radius of the Kottler hole in the expanding FLRW universe is larger when the photon exits than when it entered. In fact, it is this expansion that in their approach gives rise to the most significant part of the correction term. In particular, the authors say, without the expansion there is no $\Lambda$-correction [see after their Eq. (23)]. And later, that it arises "since the cosmological constant contributes to the extra time ... the Schwarzschild mass has to act on the passing photons." In this analysis, the geometric $\Lambda$-bending inside the vacuole of Rindler and Ishak (\cite{RindlerandIshak,IshaketalMNRAS}) goes undetected. Their lowest order correction in which $\Lambda$ appears is proportional to 
\be
m \sqrt{\Lambda},
\label{eq:Kterm}
\ee
(their equation (32)) and they find that this could cause as much as a $0.02\%$ increase in the deflection angle of the light that passes through a rich cluster. Evidently there is need to reconcile these assumptions and findings with those of previous authors. But it should also be noted that a term like (\ref(eq:Kterm)) can arise simply from aberration in an expanding universe, as we mentioned at the end of section 2.1 above.
 
Kantowski et al. in their introduction stress the desirability to produce formulas which, apart from $\Lambda$, contain only quantities that are measurably the same with and without $\Lambda$. "To conclude whether $\Lambda$ does or doesn't cause bending can easily depend on what is held in common and what property is compared in the two experiments" -- namely in a universe with and one without $\Lambda$.
%
%
\section{$\Lambda$ lensing and perturbation approaches in an FLRW model}
%
\subsection{Deflection potentials and the cosmological constant}
%
In Ref. \cite{Ishak2008}, the author shows that the contribution of the cosmological constant to the light bending angle from \cite{RindlerandIshak} can also be derived from gravitational potentials, a method that is frequently used in gravitational lensing literature, see for example \cite{Schneider1992,Bartelmann,Pyne,Carroll}. This method is of course somehow heuristic since it does not take into consideration some subtleties from the exact theory and boundary conditions. In this approach, the lens (inhomogeneity) in an FLRW background is represented by a Newtonian potential inserted  in a post-Minkowskian line element or a post-FLRW line element (see for example \cite{Schneider1992,Bartelmann,Pyne,Carroll}). The metric in such a construction is then given by $g_{ab}=\eta_{ab}+h_{ab}$ where $h_{ab}$ measure the departure from the Minkowskian metric $\eta_{ab}$.

The usual case is that of the linearized Schwarzschild spacetime; $h_{ab}$ and the associated Newtonian potential $\Phi$ can be read off the Schwarzschild metric in isotropic coordinates as (e.g. \cite{Rindler,MTW}) 
\be
ds^2=-\left(\frac{1-\frac{m}{2r}}{1+ \frac{m}{2r}}\right)^2 dt^2 + \left(1+\frac{m}{2r}\right)^4(dr^2+r^2 d\Omega^2),
\ee
and are given by 
\be
\Phi=-\frac{h_{tt}}{2}=-\frac{h_{ii}}{2}=-\frac{m}{r}
\label{eq:Schw_Pert}
\ee
Next, for a light-ray traveling in the $x$-direction, the first-order mass term of the Einstein deflection angle is then given by
\be
\alpha=-\frac{1}{2}\,\int^{+x_b}_{-x_b} \nabla_{\perp} (h_{tt}+h_{xx}) dx
2\, \int^{+x_b}_{-x_b} \nabla_{\perp} \Phi(x,y,z) dx
\label{eq:secondmethod}
\ee
where $\nabla_{\perp}\equiv\nabla-\nabla_{\parallel}$ is the gradient transverse to the path, see e.g. \cite{Schneider1992,Carroll}. At the vacuole boundary, $x_b=\sqrt{r_b^2-R^2}$ and the integration gives the usual first order result in Schwarzschild
\be
\alpha_{_{Einstein}}=\frac{4m}{R}. 
\label{eq:Einstein}
\ee
As we shall see, using this method for the Kottler metric involves some subtleties. One could argue that in linearized GR and the weak field limit, one could add the various $h_{ab}$. It therefore follows that the bending angle can be obtained by adding potentials due to different $h_{ab}$. Also, one needs to comply with the isotropic FLRW background outside the vacuole so one should also use isotropic coordinates for the de Sitter metric, see e.g. \cite{Kerr,Klioner},
\be
ds^2=-\left(\frac{1-\frac{\Lambda r^2}{12}}{1+ \frac{\Lambda r^2}{12}}\right)^2 dt^2 + \frac{1}{(1+\frac{\Lambda r^2}{12})^2}(dr^2+r^2 d\Omega^2)
\ee
which is linearized to $g_{ab}=\eta_{ab}+h_{ab}$ with, see e.g. \cite{Kerr},
\be
h_{tt}=-2\Phi = \frac{\Lambda r^2}{3},\,\,\,h_{0i}=0,\,\,\, h_{ij}=-2\Psi \delta_{ij}= -\frac{\Lambda r^2}{6}\delta_{ij}. 
\label{eq:h_Lambda}
\ee
Applying the same integration of the gradient transverse to the path gives the $\Lambda$-contribution to the bending angle as 
\be
\alpha_{\Lambda}=-\frac{1}{2}\,\int^{+x_b}_{-x_b} \nabla_{\perp} (h_{tt}+h_{xx}) dx =\, \int^{+x_b}_{-x_b} \nabla_{\perp} (\Phi+\Psi) dx = -\frac{\Lambda\,R\,r_b}{3}.
\label{eq:secondmethod}
\ee
The $\Lambda$-term (\ref{eq:secondmethod}) found using this method is equal to the $\Lambda$-term obtained from the methods discussed in  section 3.1. The method using potentials was criticized in \cite{SPH} which in turn was criticized in \cite{IRD}. In section 4.2 below it will be shown that this term can also be derived from Fermat's Principle.
%
%
\subsection{Fermat's Principle and the contribution of $\Lambda$ to the lens equation and the time delays}
%
As usual, the deflection angle can also be determined using Fermat's Principle and the Euler-Lagrange equations of the variational Principle 
\be
\delta\int n\, dl,\,\,\,\,\,where\,\,\,n= 1- 2\Phi,
\label{eq:Variation_of_N}
\ee
and $n$ is considered as an effective index of refraction of the gravitational field, $dl$ is the path of the ray of light. As discussed in for example \cite{Schneider1992}, the null curve $ds^2=g_{ab}dx^a dx^b=0$ for the Schwarzschild potential (\ref{eq:Schw_Pert}) leads to (\ref{eq:Variation_of_N}) from which the Euler-Lagrange equations give the usual result \cite{Schneider1992},
\be
\alpha_{_{Einstein}}=2 \int \nabla_{\perp} \Phi dl=\frac{4m}{R}.
\ee
In \cite{Ishak2008}, Fermat's Principle was applied in order to derive the contribution of the cosmological constant to the bending angle and use $h_{ab}$ as given by (\ref{eq:h_Lambda}). Following \cite{Ishak2008}, for a \textit{future-directed} null curve,
\be
ds^2=0
    =-dt^2\,(1-2\Phi)+(1-2\Psi)\,dl^2
    = -dt^2\,(1-\frac{\Lambda r^2}{3})+(1-\frac{\Lambda r^2}{6})\,dl^2,
\ee
or simply
\bea
dt&=&(1-(\Phi+\Psi))\,dl
  =(1+\frac{\Lambda r^2}{12})\,dl.
\label{eq:dte}
\eea
It follows from (\ref{eq:dte})  that
\be
n=1-(\Phi+\Psi)=1+\frac{\Lambda r^2}{12}.
\label{eq:n_de_Sitter}
\ee
Now, following \cite{Schneider1992}, one considers a ray of light traveling along the path $dl$ with unit tangent vector $\mathbf{e}$. The deflection angle is given by the change in the direction of the null ray. From equations (\ref{eq:Variation_of_N}) and (\ref{eq:n_de_Sitter}) and the Euler-Lagrange equations, it follows that
\bea
\frac{d\mathbf{e}}{dl}=-\left( \nabla (\Phi+\Psi)- \mathbf{e}\,(\mathbf{e}\,.\,\nabla (\Phi+\Psi)\right)=-\nabla_{\perp} (\Phi+\Psi)
\eea
and%
\bea
\mathbf{\alpha}_{\Lambda}=\mathbf{e}_{in}-\mathbf{e}_{out}=\int \nabla_{\perp} (\Phi+\Psi) dl
\eea
So for a null ray traveling in the x-direction with integration boundary ($r_b$) as given in (\ref{eq:secondmethod}), this approach yields the same $\Lambda$-term as calculated in sections 3.1 and 4.1 above, i.e. 
\be
\alpha_{\Lambda}=-\frac{\Lambda\,R\,r_b}{3}.
\ee
%
%
\subsection{Perturbations and the vacuole model}
%
Ref. \cite{SPH} considered the perturbed Robertson-Walker metric where scalar metric fluctuations are described by scalar potentials, $\Phi$ and $\Psi$:
\[
ds^2 = (1+2\Phi)\, dt^2 - a^2(t)\, (1-2\Psi)\, (d\chi^2 + \chi^2
d\Omega^2)\label{eq:FRW}
\]
and where $\chi$ is the comoving radius, $d\Omega$ is
the line element for the unit sphere, and a flat FLRW universe with no anisotropic stresses ($\Psi=\Phi$) was considered. 

In order to study the deflection inside the vacuole, the authors of \cite{SPH} aimed to find an explicit linear expression for the perturbing potentials responsible for the bending of light inside the vacuole so it can be applied in the scheme provided in section 4.1. To do that, the authors base their argument on the question of how the perturbed FLRW metric compares with the exact Kottler metric inside the vacuole. They state that the key to doing this is the transverse part of the metric. After a heuristic development, a coordinate transformation and some approximations, the authors of \cite {SPH} arrive at the following expression for their function $f$ [in their Eq. (9)]:
\bea
f = 1 -2\chi\Phi' - \dot a^2\chi^2 \left[1-2\Phi 
\left(\frac{\partial \ln |\Phi|}{\partial \ln a}+2\right)\right].
\label{eq:f_function}
\eea
The second term inside the square brackets was then dropped out under the assumption that the in a perturbed FLRW universe the Kottler vacuole around the lens is negligibly small in comparison with the Hubble length and also that, in most of the volume of the vacuole the radius $r$ is almost of the same magnitude as the proper radius of the vacuole. They compare then this expression and the Kottler function (\ref{eq:alpha}) and find that the potential $\phi$ is given by 
\be
\phi=-\frac{m}{r}-\frac{mr^2}{2R^3}+\frac{3m}{2R}
\ee
and has no $\Lambda$ term. The authors of \cite{SPH} conclude that since there is no $\Lambda$ terms appearing in their potential then there is no $\Lambda$-terms of the form $\Lambda r^2/3$ in the bending angle. They state that while there may be higher order corrections to $\phi$ involving $\Lambda$ these cannot involve a $\Lambda r^2/3$ contribution, and so they claim that a contribution to the bending angle from a $\Lambda r^2$ term derived in \cite{RindlerandIshak} and follow up work is a gauge artifact. The authors state that their solution for the potential $\phi$ is a consistant linear solution and the terms neglected were of second order.  

On the other hand, Ref. \cite{IRD} argued that the conclusion reached in \cite{SPH}, is the result of too stringent assumptions of smallness that eliminates the $\Lambda$ contribution to the bending of light by construction. Ref. \cite{IRD} argued that it was shown in \cite{IshaketalMNRAS,Ishak2008} that the radius of the vacuole, $r_b$, is much larger than the radius $R$ of the lens itself and when using any expression with the radii squared then the corresponding ratio is $(\frac{r_b}{R})^2\sim 10^3$ or larger, for example, for the cluster Abel2744 \cite{Smail1991,Allen}, this ratio is $\left(\frac{r_b}{R}\right)^2\approx 2500$, so they question this assumption. They pointed out that this is even more relevant since the bending due to the mass is achieved quickly in the vicinity of the lens and then changes very slightly as one moves away from the lens to the boundary of the vacuole, whereas the effect of $\Lambda$ inside the vacuole accumulates up to the boundary. This makes the larger size of $r_b$ significant. 
Ref. \cite{IRD} also argued against the assumption of a vacuole size negligible compared to the angular diameter distances involved in the lens equations (of Hubble scale) because the ratio of the size of the vacuole compared to the Hubble radius is in fact about the same order of magnitude ($10^{-3}-10^{-4}$) as the ratio between the $\Lambda$ contribution term to the first-order Einstein angle.
So by neglecting the size of vacuole, the effect of $\Lambda$ has been neglected with it. Mathematically, \cite{IRD} showed how not discarding the diameter of the vacuole compared to the Hubble radius affects the calculation of \cite{SPH} and restores the contribution of $\Lambda$ to the bending angle. They did not ignore the terms proportional to $\Phi$ in the square brackets of Eq. (\ref{eq:f_function}) and used the $00$-equation of the perturbed Einstein equations (see for example Eq. (5.27) in \cite{Dodelson})  
\be
k^2 \Phi +3 \frac{\dot{a}}{a}\left(\dot\Phi-\Phi \frac{\dot a}{a}\right)= 4 \pi G a^2 \rho_m \delta_m,
\label{eq:poisson}
\ee
at the same level of approximation where higher order terms on the LHS of (\ref{eq:poisson}) were kept so the approximation ${\partial \ln |\Phi|}/{\partial \ln a}=-1$ used in \cite{SPH} cannot anymore be invoked. \cite{IRD} modeled a small departure from this relation by $-1\pm \epsilon$ and followed the same procedure as in \cite{SPH} finding a potential   
\be
\Phi = -{m \over r} - {mr^2 \over 2 r_b^3} - {{\Lambda r^2}\over 12}
\label{eq:phi}
\ee
plus smaller terms. Their integration gave exactly the 2 mass terms of reference \cite{SPH} plus $\Lambda$ terms with a leading term that is precisely $-\frac{\Lambda r^2}{12}$. This $\Lambda$ term in the potential gives the same $\Lambda$ contribution to the light-bending angle (see for example \cite{Ishak2008}) as the one derived from different approaches discussed earlier.

Other different arguments where made in both \cite{SPH} and \cite{IRD} with some questions open to debate and the reader is referred to the two papers for more discussions. 
%
%
\subsection{Perturbations of the null geodesics in a first order perturbed McVittie metric}
%
Park used in \cite{Park} an approach starting with a perturbed version of an exact solution due to McVittie \cite{mcvittie} where the solution represents a mass embedded in an expanding universe. The metric is given in the notation of \cite{Park} by 
\be\label{eqn:mcvittiefull}
d s^2 = -\Big(\frac{1-\mu}{1+\mu}\Big)^2 d t^2 + (1+\mu)^4 a(t)^2 d {\bf X}^2 \,,
\ee
where $\bf X$ is the comoving coordinate, 
\be
\mu = \frac{m}{4a(t)|{\bf X} - {\bf X}_0|} \,, \nn
\ee
with $\vec X_0$ being the location of the Schwarzschild mass and the $a(t)$ the scale factor. When the mass vanishes, the metric becomes the FLRW metric and when $a(t) = 1$ the metric reduces to the Schwarzschild metric in isotropic coordinates. Park noted that in a real lensing situation in an FLRW background, the lens, the source and the observer are all moving according to Hubble's law so he proposed to define "physical" spatial coordinates by $\vec x = e^{H t} \vec X\,$, because the actual distance should be given by (scale factor)$\times$(comoving distance) where he used a scale factor $a(t) = e^{H t}$ with $H = \sqrt{\Lambda/3}$, in a universe driven by cosmological constant (a de Sitter expansion factor). Next, he noted that in weak lensing, $m$ in length units is much smaller than any other length scale under consideration, and he will work only up to ${\cal O}(m)$, and therefore he continued his work in the customized first order perturbed McVittie metric given by
{\small
\be\label{eqn:ourbkg}
d s^2 = -\Big( 1-\frac{m}{\sqrt{(x+e^{H t}q)^2 + y^2 + z^2}} \Big) d t^2 
+ \Big( 1+\frac{m}{\sqrt{(x+e^{H t}q)^2 + y^2 + z^2}} \Big) (d {\bf x} - H {\bf x} \, d t)^2 \,,
\ee
}
where he aligned the coordinates to put the lens on the $x$-axis and notes that the cosmic expansion is implemented in the coordinates themselves. The Hubble motion of lens with respect to the observer can be seen directly from the metric using $\vec x=(-e^{H t}q,0,0)$, where $q$ is a constant related to the location of the lens at a given time. 

Next, Park wrote the null geodesics equations for this first order metric, for the coordinates x and y and integrated them to first order in m and using some simplifying assumptions. He used the results to calculate, again to ${\mathcal O}(m)$, the angular location of the image of a source S observed by an observer
{\small
\bea\label{eqn:theta}
\theta &=& -\tan^{-1}\frac{y'}{x'}\Big|_{\,tO+\frac{m}{r}t_1}\\
&=& \beta + \frac{m}{\xL \ys}
\Big\{ \xs + H \rs \xL + \Big(1+H \xL\frac{\xs}{\rs} \Big)
\Big(-(1-H \rs)\xL + \sqrt{(\xs-(1-H \rs)\xL)^2+\ys^2} \;\Big) \nn\\
&& - \frac{\ys^2}{\rs^2} H \xL^2 (1-H \rs)
\log \frac{\rs^2-\xL \xs(1-H \rs)+\rs 
\sqrt{(\xs-(1-H \rs)\xL)^2+\ys^2}}{(1-H \rs)(\rs-\xs)\xL} \,\Big\} 
+ {\mathcal O}(m^2) \,,\nn
\eea
}
where $\beta = \tan^{-1} \frac{\ys}{\xs}$ is the undeflected location of the image, and $\xL=\rL/(1-H\rL)$.
Park used the approximation of a small $H$ and the identification of some expressions as being the angular diameter distances in order to re-write  the lens equation as 
\be\label{eqn:conthetaconv}
\theta = \beta + \frac{2m d_{\rm SL}}{\beta d_{\rm S} \dL}
\Big( 1 + {\cal O}(H^3) + {\cal O}(\beta^2) \Big) + {\cal O}(m^2) \,,
\ee
where 
\be
{\cal O}(\beta^2) = -\beta^2 \frac{\xs^2-4\xs\dL+2\dL^2}{4(\xs-\dL)} + \cdots\,.
\ee
Park concluded that his result is not in accord with that of \cite{RindlerandIshak} where there should be a term ${\mathcal O}(\Lambda) \sim {\mathcal O}(H^2)$.

In \cite{IRD}, the authors questioned the final result of Park since other terms including $H^2=\Lambda/3$ terms were apparently dropped out the calculation at some point, leading to the conclusion that $\Lambda$ does not contribute to lensing except via the angular diameter distances. Ref. \cite{IRD} claims that even at the same level of approximation used in Park's paper, $\Lambda$ contribution terms were simply omitted from the final lens Eq. (\ref{eqn:conthetaconv}) including a term with $H^2=\Lambda/3$, for reason of smallness. These include terms like  
\be
\frac{2md_{SL}}{d_{S}d_{L}}\, \beta\, \left(- H^2 \frac{x_S d_L^2}{2} + ... \right).
\ee 
as communicated by Park \cite{IRD}, which seems to question the conclusion reached by Park. Alternatively, \cite{SerenoII} states that the $\Lambda$-term of \cite{RindlerandIshak} is incorporated in the angular diameter distances so it can't be found separately in the treatment of Park leading to an apparent disagreement.
%
\section{Other approaches to the $\Lambda$ contribution to light bending} 
\subsection{The Weyl focusing}
%
Since some early work by R.K. Sachs \cite{Sachs1961}, it is well known that the propagation of light in spacetime can be rigorously described using the theory of geometrical optics in General Relativity \cite{Sachs1961,Dyer1977}. The distortions of light bundles are described by the optical scalar equations driven by Ricci and Weyl focusing \cite{Sachs1961,Dyer1977,Schneider1992}. The Ricci focusing is given by \cite{Sachs1961,Dyer1977,Schneider1992}
\begin{equation}
\mathcal{R}=R_{ab}k^{a}k^{b}
\end{equation} 
and the Weyl focusing (up to a phase factor) by
\begin{equation}
\mathcal{F}=C_{aibj}k^{a}k^{b}t^{i}t^{j},
\label{eq:Weyl}
\end{equation} 
where $R_{ab}$ is the Ricci tensor, $C_{aibj}$ is the Weyl curvature tensor, and $k^a$ and $t^{i}$ are null vectors and where the notation of \cite{Dyer1977,Schneider1992} was used. 

In the Kottler space ($R_{ab}=0$), so the contributions of the mass and $\Lambda$ to the bending of light come from the Weyl focusing as covariantly defined in (\ref{eq:Weyl}). Now, for the Kottler space, the non-vanishing components of the Weyl tensor are
\bea
C_{t r t r}&=&-\frac{2m}{r} \nonumber \\
C_{t \theta t \theta}&=&\frac{m}{r}(1-\frac{2m}{r}-\frac{\Lambda r^2}{3})=C_{t \phi t \phi}/\sin^2(\theta) \nonumber \\
C_{r \theta r \theta}&=&\frac{-m/r}{1-2m/r-\frac{\Lambda r^2}{3}}=C_{r \phi r \phi}/ \sin^2(\theta) \nonumber \\
C_{\theta \phi \theta \phi}&=& 2\, r\, m\, \sin^2(\theta).
\label{eq:WeylTensor}
\eea
Following the work of \cite{Dyer1977}, it is easy to verify for the Kottler metric (\ref{eq:metric}) that 
\begin{equation}
\mathcal{R}=0\,\,and\,\,\mathcal{F}=\frac{3 m h^2}{r^5} 
\label{eq:RF}
\end{equation}
(equations (27) and (35) in \cite{Dyer1977}). In Schwarzschild space, $h$ is the impact parameter but in Kottler space (not asymptotically flat) $h$ is the constant of motion $J/E$ where $J$ and $E$ are respectively the momentum and the energy of the photon. Now, we recall that the relation between the point of closest approach $r_0$ and the constant of the motion $h$ (or $b$ in other notations) is given by (e.g \cite{Wald1984,Kagramanova})
\begin{equation}
\frac{r_0^2}{h^2}=1-\frac{2m}{r_0}-\frac{\Lambda r_0^2}{3}.
\label{eq:r_0toh}
\end{equation}
For Schwarzschild, the solution $r_0(h)$ of this equation is given on page 145 of \cite{Wald1984}. This solution can be immediately generalized to include $\Lambda$ and reads 
\begin{equation}
r_0 = \frac{2}{\sqrt{\frac{3}{h^2}+\Lambda}}\cos \Big{(} \frac{1}{3} \arccos \Big{(} -3m \sqrt{\frac{3}{h^2}+\Lambda} \Big{)} \Big{)}.
\label{eq:r_0}
\end{equation}
This is in agreement with \cite{Finelli} and for $\Lambda=0$ the solution reduces to Eq. (6.3.37) in \cite{Wald1984} for Schwarzschild. 

Now, if one chooses to express the Weyl focusing fully in terms of the point of closest approach $r_0$ then (\ref{eq:r_0toh}) and (\ref{eq:RF}) yield  
\begin{equation}
\mathcal{F}=\frac{3 m h^2}{r_0^5} = \frac{3 m}{r_0^3} \Big{(} 1- \frac{2m}{r_0}- \frac {\Lambda r_0^2}{3}  \Big{)}^{-1}. 
\end{equation}
So $\Lambda$ is present in the final result. Similarly, if one chooses to express the focusing fully in terms of $h$, then using (\ref{eq:r_0}) in order to eliminate $r_0$ from (\ref{eq:RF}) gives the Weyl focusing, $\mathcal{F}$, as a function of $m$, $h$, and $\Lambda$. In a way or the other the final expression for the Weyl focusing has a $\Lambda$ in it. 

This general result is independent of any angle calculations and supports independently the conclusion that $\Lambda$ does contribute to the bending of light in the Kottler space in an invariant and thus coordinate-independent way. 
%
%
\subsection{Using Friedmann-Robertson-Walker coordinates}
%
In Ref. \cite{Khriplovich}, the authors considered an invariant built out of the light frequency and the small angle between two incident rays that can both in principle be measured at the point of observation. They compared this invariant in the de Sitter coordinates and the Friedmann-Robertson-Walker coordinates arriving to the conclusion that the $\Lambda$ light-bending terms drop out of the lensing equation. However, the authors state in the penultimate paragraph of their paper the following: ``Let us note that some corrections on the order of $\lambda^2 \rho^2\sim \Lambda r_r r_o$ to the lensing effects may exist, as well as other cosmological corrections in the general case of the FLRW Universe. However, such ``short-distance" phenomena are perhaps too small to be of practical interest." 

It was argued in \cite{IRD} that \cite{Khriplovich} did an expansion of their Eq. (16) keeping only the first term in the invariant considered there and dropping the second term of that equation, i.e. $\omega \lambda \rho$ where $\lambda\equiv\sqrt{\Lambda/3}$. The neglected term in question is $\Lambda r_g r_o/3$ and represents a contribution of $\Lambda$ to the squared invariant that was discarded by assumption. 
%
%
\section{Concluding remarks} 
%
As initiated in the original paper of this series \cite{RindlerandIshak} 
and supported by several other papers \cite{SerenoI,Lake,SerenoII,SchuckerI,SchuckerII,SchuckerIII}, although the cosmological constant drops out of 
the coordinate orbital equation of light rays in Kottler spacetime, it does 
nevertheless contribute to the bending of light through the metric itself which
determines the geometric and measurable properties of the coordinate orbital
equation. This corrects a long-standing misconception that the cosmological
constant does not affect the observed light bending angle. Some references
\cite{SerenoI,SerenoII,SchuckerIII} suggest that there can be an apparent 
disagreement between some papers due to the interpretion of the $\Lambda$ lensing terms. 
For example \cite{SerenoI,SerenoII} assert that the $\Lambda$ term discussed in 
the above sections can be viewed either as an additional term to the bending 
angle or can be incorporated into the angular diameter distances. \cite{SerenoII} 
also states that this is the reason why the $\Lambda$-term was not found separately 
in \cite{Park,Khriplovich}. Other papers \cite{IRD} have argued that $\Lambda$-terms 
were missed because of too stringent assumptions of smallness and were neglected 
in the perturbation expansions. In sum, it seems fair to assert that the fact that 
$\Lambda$ affects the bending of light has by now been well established, while 
putting the $\Lambda$ light-bending terms into a cosmological context is still 
subject to some interpretation and requires further work and clarification.
%
%
%

\begin{acknowledgements}
This material is based upon work supported in part by NASA under grant NNX09AJ55G.
\end{acknowledgements}


\begin{thebibliography}{}
%
\bibitem{Eddington} Eddington A. S., The Mathematical Theory of Relativity, The University Press, Cambridge (1923). 
%
\bibitem{Rindler1} Rindler W., Essential Relativity, 1st ed., page 208 (1969). 
%
\bibitem{Islam} N.J. Islam, Phys. Lett. A  {\textbf{97}}, 239 (1983).
%
\bibitem{Kottler} F. Kottler, Ann. Phys. {\textbf{361}}, 401 (1918).
%
\bibitem{RindlerandIshak} W. Rindler, M. Ishak, Phys. Rev. D \textbf{76} 043006 (2007).
%
\bibitem{Bonnor} Bonnor W.B., in Recent Developments in General Relativity, Pergamon, New York (1962).
%
\bibitem{Flamm} L. Flamm, Physik. Z. {\textbf{17}}, 448 (1916). 

\bibitem{Rindler} Rindler W., Relativity: Special, General, and Cosmological, Second Edition. Oxford University Press, (2006).

\bibitem{IRD}Ishak M., Rindler W., Dossett J.,  Mon. Not. Roy. Astron. Soc. {\textbf{403}}, 2152-2156 (2010).

\bibitem{Bodenner} Bodenner J., Will C., Am. J. Phys., Vol. {\textbf{71}}, No. 8, 770 (2003).

\bibitem{IshaketalMNRAS} Ishak M., W. Rindler, J. Dossett, J. Moldenhauer, C. Allison, Monthly Notices of the Royal Astronomical Society, Volume \textbf{388}, Issue 3, p1279 (2008).

\bibitem{MTW} C.W. Misner, K.S. Thorne and J.A. Wheeler,
  \textit{Gravitation} (Freeman, San Francisco 1973).

\bibitem{Keeton} Keeton C., Petters A., Phys. Rev. D{\textbf{72}}, 104006 (2005).

\bibitem{SerenoI} Sereno M., Phys.Rev. D {\textbf{77}}, 043004 (2008).

\bibitem{SerenoII} Sereno M., Phys. Rev. Lett. {\textbf{102}}, 021301 (2009). 

\bibitem{SchuckerI} Schucker T., General Relativity and Gravitation, \textbf{41}:67-75,(2009). 

\bibitem{Sharon} Sharon K. et al., Astrophys. Jour. 629, L73 (2005); N. Ota, et al., Astrophys. Jour., 647, 215 (2006).

\bibitem{SchuckerII} Schucker T., N. Zaimen, Astrono. \& Astrophys., \textbf{484}, 103 (2008).

\bibitem{SchuckerIII} Schucker T., General Relativity and Gravitation \textbf{41}:1595-1610  (2009).  


\bibitem{Bartelmann}Bartelmann M., Schneider P. Phys.Rept. 340, 291 (2001).

\bibitem{Israel} Israel W., Nuovo Cim B {\bf 44}, 1. (1966). Erratum {\bf 48}, 463. 

\bibitem{Darmois} Darmois G.{\em M\'{e}morial de Sciences Math\'{e}matiques, Fascicule XXV}, ``Les equations de la gravitation einsteinienne'', Chapitre V. (1927).

\bibitem{Swiss-cheese} Einstein A. and Strauss E., Rev. Mod. Phys. \textbf{17}, 120.(1945), erratum: ibid {\bf 18}, 148 (1946). 

\bibitem{Swiss-cheese2} Schucking E., Z. Phys. 137, 595 (1954); and a long list of references in \cite{Krasinski} 

\bibitem{Krasinski} Krasinski A., \textit{Inhomogeneous Cosmological Models} (Cambridge University Press, Cambridge, 1997). 

\bibitem{RindlerLambda} Rindler W., Astrophys. J. \textbf{157}, L147 (1969). 

\bibitem{obs1} Riess A., {\em{et al.}}, Astron. J. {\textbf{116}}, 1009-1038 (1998).
\bibitem{obs2} Perlmutter S., {\em{et al.}}, Astrophys. J. {\textbf{517}}, 565-586 (1999).
\bibitem{obs3} Knop R., {\em{et al.}}, Astrophys. J. {\textbf{598}}, 102-137 (2003).
\bibitem{obs4} Riess A., {\em{et al.}}, Astrophys. J. {\textbf{607}}, 665-687 (2004).
\bibitem{obs5} Bennett C., {\em{et al.}}, Astrophys. J. Suppl. Ser. {\textbf{148}}, 1 (2003).
\bibitem{obs6} Spergel D., {\em{et al.}}, Astrophys. J. Suppl. Ser. , 175 (2003).
\bibitem{obs7} Page L. et al., Astrophys. J. Suppl. Ser. {\textbf{148}}, 2333 (2003).
\bibitem{obs8} Seljak U. et al., Phys.Rev. D{\textbf{71}}, 103515 (2005).
\bibitem{obs9} Tegmark M., {\em{et al.}},  Astrophys. J. {\textbf{606}}, 702-740 (2004).
\bibitem{obs10} Spergel D., {\em{et al.}}, Astrophys. J. Suppl. 170, 377 (2007).

\bibitem{Sereno} Sereno M., Jetzer Ph., Phys. Rev. D {\textbf{73}}, 063004 (2006).

\bibitem{Kagramanova}  Kagramanova V., Kunz J., Lammerzahl C., Phys.Lett. B {\textbf{634}} 465-470 (2006).

\bibitem{Kantowski} Kantowski R., Chen B., Dai X, arXiv:0909.3308  (2009), ApJ submitted.

\bibitem{Allen} Allen S., M.N.R.A.S, \textbf{296}, 392, (1998).

\bibitem{Smail1991} Smail I. et al., M.N.R.A.S, 252, 19 (1991).

\bibitem{Ishak2008} Ishak M., Phys. Rev. D \textbf{78}, 103006, (2008).

\bibitem{Carroll} Carroll S., \textit{Spacetime and Geometry: An Introduction to General Relativity} (Addison Wesley, San Fransisco, 2004).

\bibitem{Pyne} T. Pyne, M. Birkinshaw Astrophys.J. 458 (1996) 46. 

\bibitem{Schneider1992} P. Schneider, J. Ehlers, E.E. Falco, \textit{Gravitational Lenses} Springer-Verlag (1992). 

\bibitem{Kerr} A. W. Kerr, J. C. Hauck, B. Mashhoon,  Class. Quant. Grav., {\textbf{20}}, 2727 (2003).

\bibitem{Klioner}
S. Klioner, M. Soffel, Proceedings of the Symposium "The Three-Dimensional Universe with Gaia", 2004, Observatoire de Paris-Meudon, France (ESA SP-576). 

\bibitem{SPH} Simpson F., J. A. Peacock, A. F. Heavens,
Mon. Not. Roy. Astron. Soc. {\textbf{402}}, 2009-2016 (2009).

\bibitem{Dodelson} Dodelson S., {\textit{Modern Cosmology}}, Academic Press (2003).

\bibitem{Park} Park M.,   Phys. Rev. D.\textbf{78}, 023014 (2008).  

\bibitem{mcvittie} McVittie G. C., Mon. Not. R. Astron. Soc. {\bf 93} 325 (1933).

\bibitem{Lake} Lake K., arXiv:0711.0673.

\bibitem{Sachs1961}Sachs R.K., Proc. Roy. Soc. London. A {\textbf{264}}, 309 (1961).

\bibitem{Dyer1977} Dyer C., Mon. Not. Roy. Astro. Soc. {\textbf{180}}, 231 (1977).

\bibitem{Wald1984} R. Wald \textit{General Relativity} (University of Chicago Press, 1984).

\bibitem{Finelli} F. Finelli, M. Galaverni, A. Gruppuso, Phys.Rev. D {\textbf{75}}, 043003 (2007).

\bibitem{Khriplovich} Khriplovich I., A. Pomeransky, Int. J. Mod. Phys. D \textbf{17}, 2255-2259 (2008).


\end{thebibliography}
%
%

\end{document}